\documentclass[aps, pra,twocolumn,longbibliography, floatfix]{revtex4-1}

\usepackage{amsmath,amssymb}
\usepackage{graphicx,color}
\usepackage{hyperref}
\usepackage{bm}
\usepackage{epstopdf}
\usepackage{dcolumn}   
\usepackage{array,multirow,graphicx}

\begin{document}

\title{Long-range connectivity in a superconducting quantum processor using a ring resonator} 
\author{Sumeru Hazra,$^{1}$ Anirban Bhattacharjee,$^{1}$ Madhavi Chand,$^{1}$ Kishor V. Salunkhe,$^{1}$}
\author{Sriram Gopalakrishnan}
\altaffiliation{Current affiliation: Institute of Quantum Computing, University of Waterloo, Waterloo, Ontario N2L 3G1, Canada}
\author{Meghan P. Patankar,$^{1}$}
\author{R. Vijay$^{1}$}
\email{Corresponding author: r.vijay@tifr.res.in}
\affiliation{$^{1}$Department of Condensed Matter Physics and Materials Science,Tata Institute of Fundamental Research, 1 Homi Bhabha Road, Mumbai 400005, India}

\date{\today}

\begin{abstract}
	Qubit coherence and gate fidelity are typically considered the two most important metrics for characterizing a quantum processor. An equally important metric is inter-qubit connectivity as it minimizes gate count and allows implementing algorithms efficiently with reduced error. However, inter-qubit connectivity in superconducting processors tends to be limited to nearest neighbour due to practical constraints in the physical realization. Here, we introduce a novel superconducting architecture that uses a ring resonator as a multi-path coupling element with the qubits uniformly distributed throughout its circumference. Our planar design provides significant enhancement in connectivity over state of the art superconducting processors without any additional fabrication complexity. We theoretically analyse the qubit connectivity  and  experimentally  verify  it  in a device capable of supporting up to twelve qubits where each qubit can be connected to nine other qubits.  Our concept is scalable, adaptable to other platforms and has the potential to significantly accelerate progress in quantum computing, annealing, simulations and error correction.

\end{abstract}

\maketitle 

\section{Introduction}
Quantum information processors promise tremendous computational advantages in solving a broad class of problems, offering polynomial\cite{grover} and even sometimes exponential\cite{shor} speed ups compared to the best known implementation in classical computers. Continual efforts across several platforms have propelled the field from proof of concept demonstrations with a few qubits to actual processors with tens of qubits\cite{ibmqv,51q_iontrap,16q_alltoall,rydberg}, leading to the recent milestone of achieving quantum supremacy\cite{quantum_supremacy}. 
Quantum algorithms generally assume the ability to implement gates between any arbitrary pair of qubits in a processor. However, it is often impractical or topologically impossible to engineer a processor with arbitrarily long-range coupling. Therefore, in order to implement an arbitrary quantum operation in such a constrained architecture, the quantum data needs to be transported across several connected qubits in the grid, leading to additional operations and consequently more errors. More recently, the importance of qubit connectivity for quantum annealers\cite{annealing_floquet} and the performance of near term quantum processors as quantified by the quantum volume metric\cite{quantum_vol} have also been studied. 

\begin{figure*}[t]
	\centering
	\includegraphics[width=0.95 \textwidth]{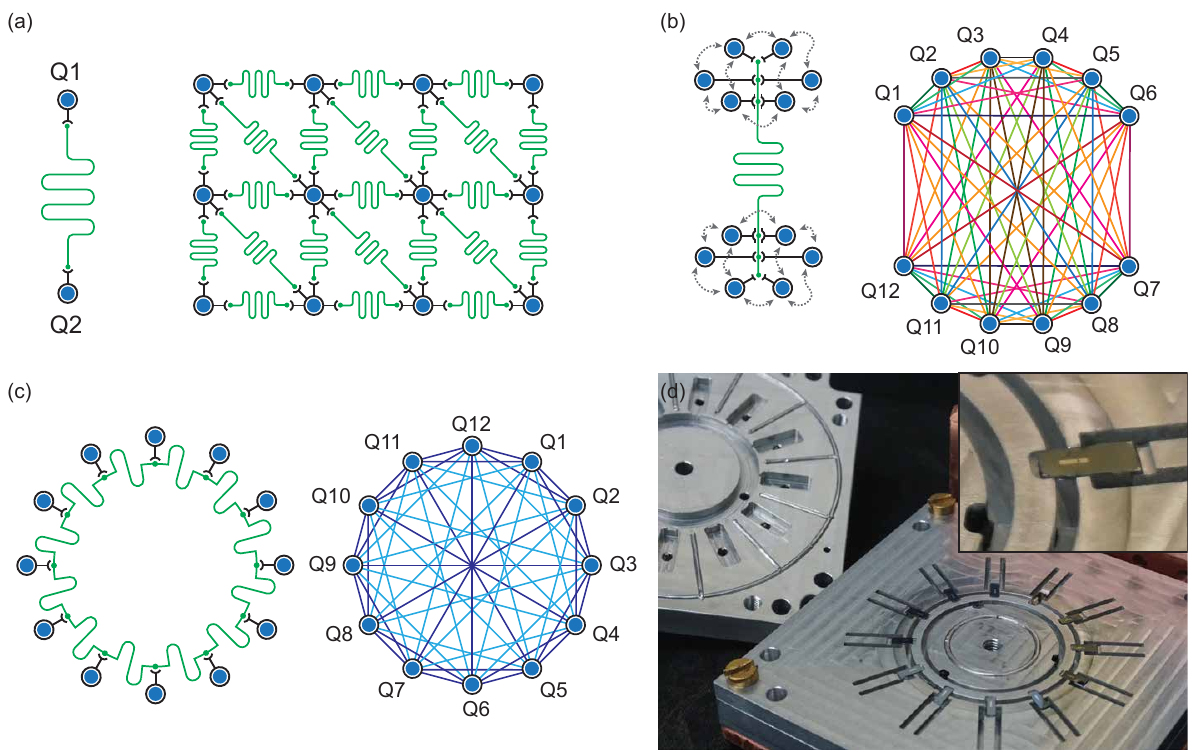} 
	\vspace{-5pt}
	\caption{Inter-qubit connectivity in superconducting circuit architecture. (a) A standard bus resonator connecting two qubits positioned at its anti-nodes. Extension to a 2D grid layout which achieves uniform but only nearest neighbor coupling.  
	(b) Several qubits placed in close proximity at each anti-node of a bus resonator produces all-to-all connectivity. However, one typically observes wide variation in inter-qubit coupling and unwanted cross-talk (grey dashed arrows)
	(c) Our design uses a ring resonator as a multi-path coupling element connecting several qubits. It offers a highly connected scalable qubit network with only two slightly different values for inter-qubit coupling with negligible cross-talk.
	(d) Image of a semi-assembled 3D cQED implementation of a 12-qubit network. The qubit slots (magnified in the inset) are placed $30^o$ apart and the readout resonators are $\lambda/4$ sections of transmission line extending radially outward.
	}
	\label{fig:outline}
\end{figure*}

Among the two leading platforms for a practical quantum processor, all-to-all connectivity has been demonstrated in the ion-trap system for up to eleven qubits\cite{ion_11} whereas even the most powerful superconducting processors today provide only nearest neighbour connectivity achieved by direct capacitive coupling\cite{surface}, tunable couplers\cite{google_chip} or bus resonators\cite{quantum_bus}. Fig.~\ref{fig:outline}(a) shows a typical planar layout of qubits connected by bus resonators. While recent experiments\cite{atomic_cat,16q_alltoall} have implemented all-to-all connectivity by using a single bus resonator with many superconducting qubits connected at its anti-nodes (Fig.~\ref{fig:outline}(b)), the inter-qubit coupling strength varies significantly from pair to pair and the close proximity of qubits leads to unwanted cross-talk.

In this article, we propose an alternative scalable architecture for a dense network of superconducting qubits with significantly enhanced connectivity. The key idea in this architecture is the use of a ring resonator as a bus cavity capable of mediating interaction between several qubits. The qubits are distributed throughout the circumference of the ring resonator and thus spatially separated. This provides good microwave isolation resulting in individual qubit addressability and negligible qubit cross-talk\cite{atomic_cat}. The architecture proposed here is completely agnostic to the kind of two-qubit gates to be implemented and is compatible with both flux activated gates with tunable qubits\cite{surface,fast_flux_jm,schuster_tunable,dicarlo_tunable,rigetti_para} and microwave drive activated gates with fixed frequency qubits\cite{cr_rigetti,raman_eth} offering wide compatibility. Further, this idea can be extended to other architectures which rely on bus cavities for coupling \cite{jesonpetta-princeton,magnon}.

\section{Ring Resonator Coupler: Theory}
\begin{figure}[t]
	\centering
	\includegraphics[width=0.48 \textwidth]{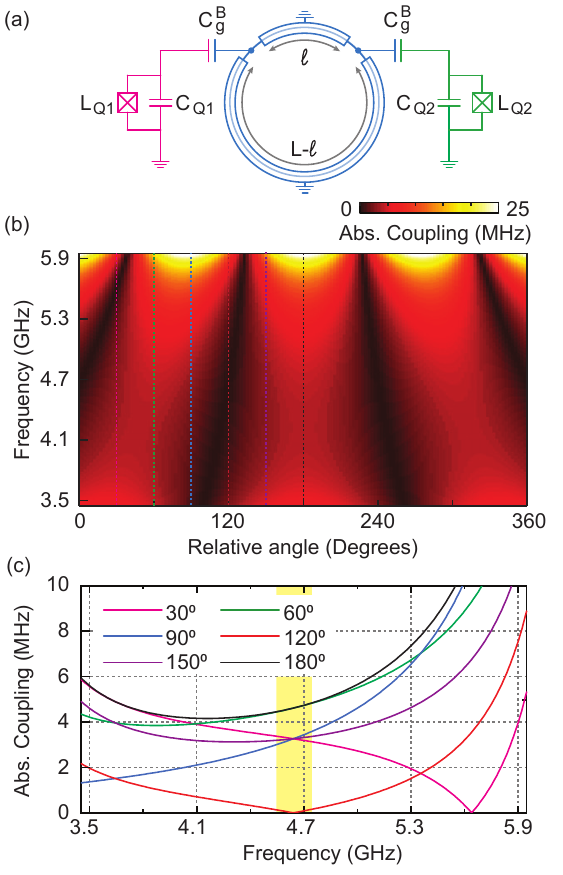} 
	\vspace{-15pt}
	\caption{(a) Circuit schematic showing a pair of qubits (magenta and green) connected to the ring resonator (blue) with the fundamental mode at 3.1 GHz. Inter-qubit angle $\theta=2\pi\ell/L$ determines the coupling between the pair of qubits at a particular frequency. 
    (b) Numerical estimation of inter-qubit coupling as a function of qubit frequency and the angular separation between two identical qubits. 
	The qubits are operated near the mean of the two modes of the bus resonator. 
	(c) Vertical line cuts of plot B for six specific angles spaced $30^o$ apart along the ring, showing inter-qubit coupling as a function of qubit frequency. Ideal range of operating frequencies for qubits with minimal coupling variation is highlighted in yellow.}
	\label{fig:theory}
\end{figure}

The working principle of the ring resonator coupler is based on wave interference between two independent paths connecting any pair of qubits\cite{giant_atom_theory,giant_atom}. We start by considering only two transmon\cite{transmon} qubits in a ring of circumference $L$ and calculate the exchange coupling using a linearized model\cite{blackbox_quant_divincenzo,blackbox_quant_girvin}. 
We place the two qubits at an arbitrary angle $\theta$ with respect to each other so that they are connected by two transmission lines of characteristic impedance $Z_R$ and lengths $\ell=(\theta/{2\pi})L$ and $\ell '=(1-(\theta/{2\pi}))L$ as shown in Fig. \ref{fig:theory}(a). We denote the modes of the ring resonator by $\left\lbrace \omega_R^0, \omega_R^1, ..., \omega_R^n, ...\right\rbrace $ and the qubit frequencies are chosen between the first two modes of the ring resonator.

 We compute the inter-qubit coupling $J_{ij}$ (see Appendix A for details) and plot it as a function of inter-qubit angle and qubit frequencies $\omega_Q$ in Fig. \ref{fig:theory}(b). We observe that the inter-qubit coupling is reasonably flat except near angles $120^o$ and $240^o$ and frequencies close to the resonant modes of the ring resonator.  If we choose $30^o$ angular spacing between qubits and operate around frequency $\tilde{\omega}_Q=\left(\omega_R^0+\omega_R^1\right)/2$, connected pairs have only two different values of finite coupling. The vertical line cuts in Fig. \ref{fig:theory}(b) are separately plotted in Fig. \ref{fig:theory}(c) to highlight the coupling as a function of qubit frequencies for these angles. Qubit pairs placed at $60^o$, $180^o$ and $300^o$ show a maximum value of coupling while those placed at $30^o$, $90^o$, $150^o$, $210^o$, $270^o$ and $330^o$ show a coupling reduced by a factor of $\sqrt{2}$ compared to the maximum value. However, the qubits placed at $120^o$ and $240^o$ show zero coupling at this special frequency $\tilde{\omega}_Q$ due to destructive interference of the two paths connecting the two qubits\cite{giant_atom_theory}. Thus in a qubit network realized with this particular geometry, each of the twelve qubits is coupled to nine other qubits (Fig.~\ref{fig:outline}(c)). Further, the inter-qubit coupling is a slowly varying function of qubit frequency around $\tilde{\omega}_Q$ and thus allows us enough flexibility in the choice of qubit frequencies with a nominal variation in the inter-qubit coupling (less than $\pm10\%$ in a range of $\pm 100$ MHz; the region highlighted in yellow in Fig. \ref{fig:theory}(c)). 

\section{Experimental demonstration of inter-qubit coupling}
\begin{figure}[t]
	\centering
	\includegraphics[width=0.48 \textwidth]{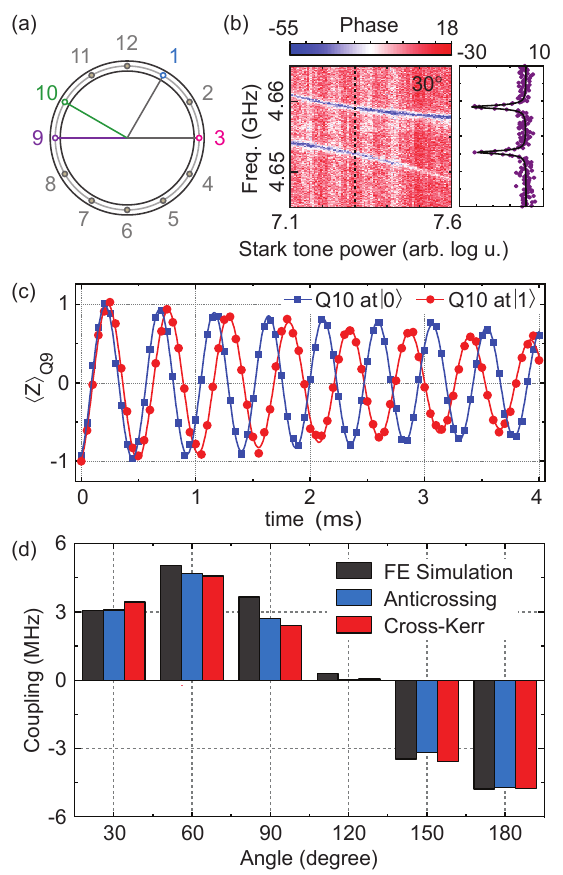} 
	\vspace{-15pt}
	\caption{Measurement of inter-qubit coupling. (a) The relative locations of the four qubits coupled with the ring resonator that allow measuring all six possible combinations in the current set-up. (b) An anti-crossing between the pair Q9 and Q10 observed while Stark shifting the transmon Q10 and performing a spectroscopy on Q9. A vertical line cut showing the hybridized states of two-qubits in the right panel when they are brought on resonance. (c) A set of conditional Ramsey fringes obtained for the pair Q9 and Q10 in the  conditional Ramsey experiment. The measured cross-Kerr shift is used to estimate the inter-qubit coupling.(d) Inter-qubit couplings between all possible pairs estimated from the finite element simulation of the structure and experimentally measured using anti-crossing and by conditional Ramsey experiment.}
	\label{fig:exp_main}
\end{figure}

In our experiment, we use the 3D circuit QED architecture consisting of a central circular bus resonator realized in rectangular coax geometry with twelve slots distributed along its perimeter at $30^o$ angular spacing where the qubit chips can be placed. The symmetry of the design ensures that only four qubits are sufficient to explore all six possible combinations of inter-qubit angles in this design. These four qubits are placed at positions 1, 3, 9 and 10, as shown in Fig. \ref{fig:exp_main}(a) and we label them as Q1, Q3, Q9 and Q10 respectively. The remaining qubit positions are filled with dummy silicon chips to retain the symmetry of the ring resonator modes. Details of the cavity design and assembly are discussed in Appendix C. The fundamental mode and the first harmonic of the bus cavity are measured at 3.127 GHz and 6.240 GHz in a separate experiment. All device parameters and coherence properties extracted from the experiments are listed in Table \ref{table:deviceparameters} in Appendix C.

We measure the vacuum Rabi splitting between a pair of qubits by  tuning them into resonance. All qubits are tuned by applying an off-resonant  ac Stark shift tone detuned by $\Delta_S/2\pi \simeq 250$ MHz from the qubit's original transition frequency. We perform spectroscopy on one of the qubits while varying the amplitude of the Stark shift tone to obtain the vacuum Rabi splitting spectrum of the hybridized qubits (see Fig. \ref{fig:exp_main}(b)). The minimum vacuum Rabi splitting on each pair is measured while parking the spectator qubits at particular frequencies away from the crossing point. Finally, we estimate the transverse coupling strengths $J_{ij}$, that provide the best match with the experimentally observed splittings by numerically solving the following Hamiltonian for the 4-qubit system
  \begin{equation}
\label{hm}
\begin{split}
\hat{\mathcal{H}}_0/\hbar=&\sum_{i=1}^{4}(\omega_i\hat{a}^\dagger_i\hat{a}_i+\frac{\delta_i}{2}\hat{a}^\dagger_i\hat{a}_i(\hat{a}^\dagger_i\hat{a}_i-\mathbf{I}))\\&+\sum_{i<j,j=1}^{4}J_{ij}(\hat{a}^\dagger_i\hat{a}_j+\hat{a}^\dagger_j\hat{a}_i).
\end{split}
\end{equation}
Here $\omega_i$ and $\delta_i$ are the transition frequency and anharmonicity of the $i^{th}$ transmon respectively and $J_{ij}$ is the pairwise exchange coupling between $i^{th}$ and $j^{th}$ transmon. See Appendix E for more details.

We perform another experiment which uses the cross-Kerr effect between pairs of qubits to estimate $J_{ij}$ and obtain similar results. The pair-wise cross-Kerr shift is measured by using a conditional Ramsey sequence on one qubit while keeping the other qubit in $|0\rangle$ and $|1\rangle$ respectively as shown in Fig. \ref{fig:exp_main}(c). The extracted values of $J_{ij}$ obtained for each pair from the two experiments along with the results from finite element simulation are plotted in Fig. \ref{fig:exp_main}(d) (the raw data for all pairs are shown in the Supplemental Materials\cite{supp}). The values exhibit good agreement with the theoretical model for the given frequency range, confirming highest coupling strengths for pairs at $60^o$ and $180^o$, intermediate coupling strengths for $30^o$, $90^o$ and $150^o$ and a negligible coupling for the pair at $120^o$. We attribute the slight deviation from the ideal values of coupling to variability in the qubit chip dimensions and misalignment of qubit chips while positioning it in the slots. 
In addition the cross-Kerr effect is sensitive to the ring resonator response at qubit transition frequencies from $|1\rangle$ to $|2\rangle$ as well and that effect is not included in our theory (See Appendix D for details). We also characterize the microwave cross-talk between the qubits at readout frequencies as well as at qubit frequencies and observe that the physical separation of the qubits in this architecture along with 3D microwave enclosure leads to very low classical cross-talk\cite{hamiltonian_tomography_ibm} in our set up. Details of the cross-talk measurements are given in Appendix F and G.


\section{Scaling up to larger processor}
The unique ability to couple several non-nearest neighbor qubits demonstrated in this architecture readily suggests multiple potential extensions to achieve highly connected quantum processors. Simply putting more qubits in the existing design (with reduced angular spacing) will introduce a wider variation of coupling. Instead, we propose two extensions which also provide only two values of inter-qubit coupling between different pairs. Further, we suggest using a tunable coupler between each qubit and the ring resonator to turn on couplings only between the intended qubits while keeping the other qubits completely isolated. This will provide maximum flexibility for different applications. 

\begin{figure*}[!t]
	\centering
	\includegraphics[width=0.95\textwidth]{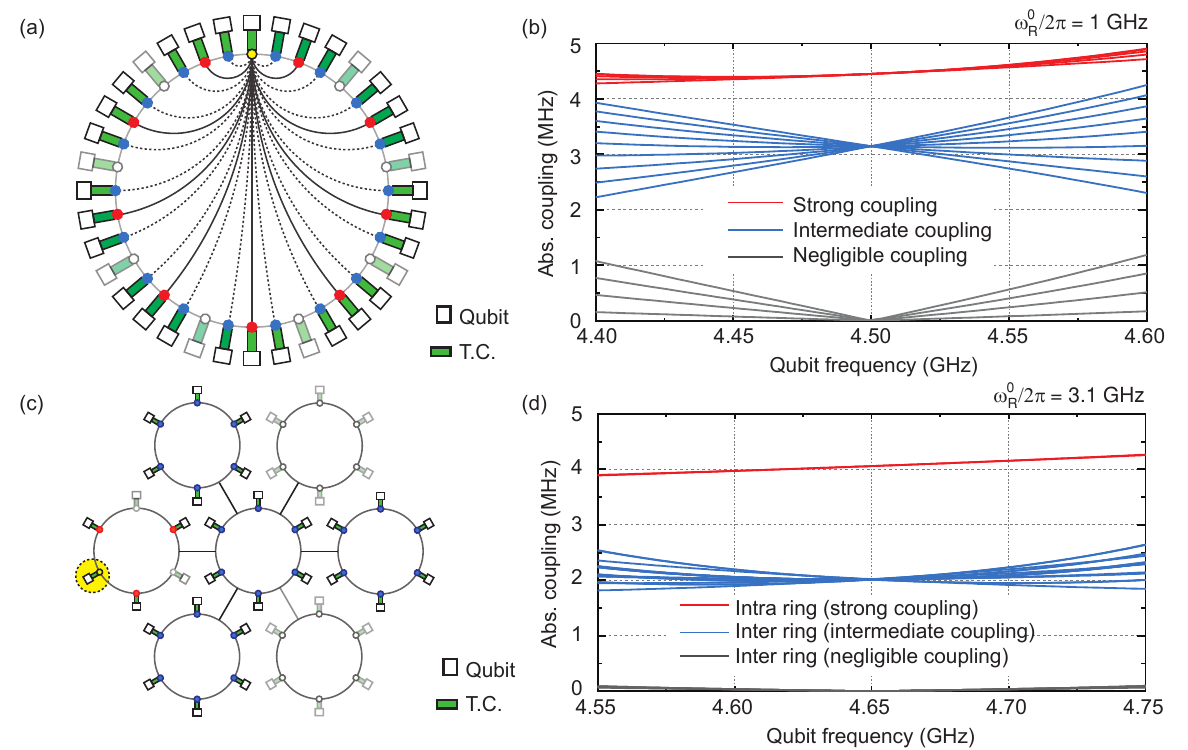} 
	\vspace{-5pt}
	\caption{Scaling to larger number of qubits. (a) A larger ring with the fundamental mode at $\omega_R^0/2\pi= 1$ GHz, accommodating 36 qubits. The qubit frequencies are chosen around the special frequency at the mean of two higher harmonics $\omega_R^3$ and $\omega_R^4$ of the ring resonator. A tunable coupler is suggested to control the coupling of each qubit to the network. (b) Variation of inter-qubit coupling around the special point for the larger ring architecture. Each qubit is connected to 27 other qubits in the ring with only two different values of couplings shown in red and blue at the special frequency. (c) A multi-ring architecture with 42 qubits placed at the marked positions. Each of the rings has a fundamental mode at $\omega_R^0/2\pi=3.1$ GHz with the operating frequency at 4.65 GHz. Two rings are connected via a $\lambda/2$ section of transmission line with a characteristic impedance equal to half that of the ring resonators. (d) Variation of coupling for different set of qubit pairs marked in (c). All the qubits placed in the position marked with blue share identical coupling at the special frequency whereas the positions marked with red share a higher coupling. The faded out qubits are not coupled to the highlighted qubit.}
	\label{fig:extension}
\end{figure*}

The simplest extension is to increase the length of the ring resonator e.g. using one with the fundamental mode $\omega_R^0=1$ GHz which can contain 36 qubits with $10^o$ angular spacing. The qubits are now operated between the third and fourth harmonics of the ring resonator around the mean frequency $\tilde{\omega}_Q=(\omega_R^3+\omega_R^4)/2=4.5$ GHz. In this geometry every qubit is connected to 27 other qubits in the network with only two different values of coupling as shown by the solid and dotted connections in Fig. \ref{fig:extension}(a). The dependence of the coupling on frequency around the special point is numerically calculated and plotted in Fig.  \ref{fig:extension}(b). However, as shown in the figure, the frequency dependence of the coupling grows stronger as the fundamental frequency of the ring resonator is pushed down to accommodate a larger number of qubits.

The second extension uses multiple identical ring resonators containing six qubits each and connected to each other by $\lambda/2$ sections of transmission line resonating at the fundamental mode of the ring resonator (See Fig. \ref{fig:extension}(c)). The connecting $\lambda/2$ section has a characteristic impedance half that of the ring resonators. In this design with 42 qubits, each qubit in the outer ring is connected to 27 other qubits whereas each qubit in the central ring is connected to 39 other qubits. We numerically calculate the frequency dependence of the coupling for different qubit locations and plot them in Fig. \ref{fig:extension}(d), showing two different coupling values for inter-ring and intra-ring qubits respectively. The detailed calculations related to these two extensions are discussed in Appendix B.

\section{Conclusion}
In summary, our results demonstrate a powerful coupling architecture using a ring resonator to realize a highly connected qubit network for superconducting circuits. The unique feature offered by this design can be easily translated to a wide range of qubits\cite{transmon,csfq,fluxonium}, implemented in both 2D and 3D layout and adapted to any platform which relies on a cavity bus for mediating inter-qubit coupling. This will substantially enhance the performance of present day quantum processors without any added topological complexity or control wiring overhead. We anticipate that the enhanced qubit connectivity will have a significant impact in the field of quantum simulations\cite{fnori_qsim} and error correction\cite{campbell_qec}.

\begin{acknowledgments}
We acknowledge the nanofabrication and central workshop facilities of TIFR. We thank Mandar Deshmukh, Michel Devoret and Kater Murch for useful inputs on the manuscript. This work was supported by the Department of Atomic Energy of the Government of India. We also acknowledge support from the Department of Science and Technology, India, via the QuEST program.
\end{acknowledgments}
{
\appendix
\section{Even and odd mode analysis of the ring resonator}
In order to theoretically investigate the coupling between any arbitrary pair of qubits connected to the ring at some particular angle, we first extract the ABCD matrix of the network between any two points of the ring resonator, shown in the dashed box in Fig \ref{fig:even_odd}(a).  We use this symmetry of the system to decompose it into a superposition of two simpler circuits, as shown in Fig. \ref{fig:even_odd}(b). We define two different modes of excitation for the circuit\cite{pozar}; the even mode where $V_{g1}=V_{g2}=V/2$, and the odd mode, where $V_{g1}=-V_{g2}=V/2$. Now, from the superposition of these two modes we get an excitation $V_{g1}=V$ and $V_{g2}=0$. We define the impedance of the ring cavity as $Z$ and now treat the two modes separately.

For the even mode of excitation, $V_{g1}=V_{g2}=V/2$, so there is no current flowing through the arms of the resonator. Therefore we can bisect the ring along the vertical symmetry axis with open circuits at the points of bisection. Similarly, for the odd mode of excitation, $V_{g1}=-V_{g2}=V/2$, and hence there is a voltage null at the points of bisection. Therefore we can ground the central pin at these two points of the resonator, leading to short circuit. 

\begin{figure}[t]
	\centering
	\includegraphics[width=0.48 \textwidth]{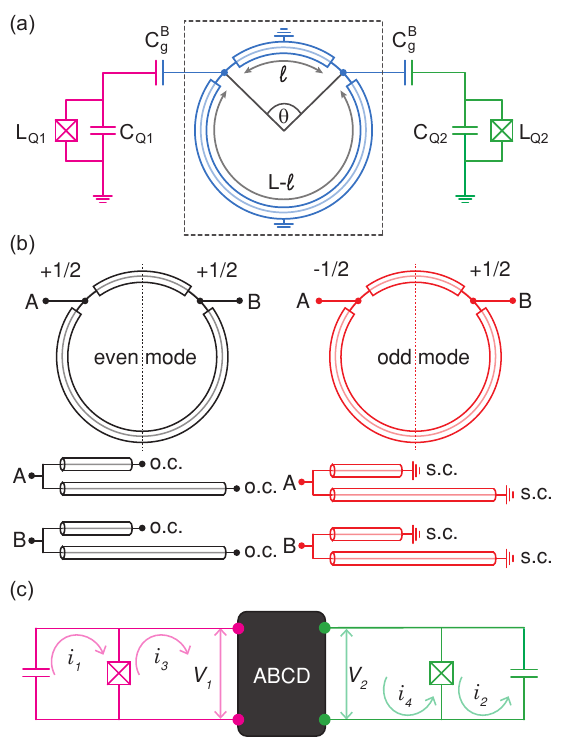} 
	\vspace{-10pt}
	\caption{(a) Circuit schematic showing a pair of qubits (magenta and green) connected to the ring resonator (blue). Qubits are operated around the mean frequency of the first two modes of the ring resonator. (b) Breaking down the ring resonator into a superposition of two simpler circuits named even and odd modes. (c) Solving the coupled qubit system using the ring resonator as the coupling element. The ring is treated as a two port black box with given network parameters. The junctions are treated as linear inductors.}
	\label{fig:even_odd}
\end{figure}

As shown in Fig. \ref{fig:even_odd}(a), the two qubits are connected to each other by two sections of transmission line of length $\ell=L(\theta/\pi)$ and $L-l$, where $L$ is the total circumference of the ring and $\theta$ is the angular separation between the two qubits in radians. We use the impedance transformation formula to get the effective impedance of the short and open circuit at the end of a transmission line of characteristic impedance $Z_R$ and length $x$.
\begin{equation}
\label{impedance_transform}
Z_{\rm{in}}=Z_R\frac{Z_L+jZ_R\tan\beta x}{Z_R+jZ_L\tan\beta x}
\end{equation}
where, $\beta=2\pi/\lambda$ and the load impedance, $Z_L$ is equal to $0$ and $\infty$ respectively for the short circuit (odd) and the open circuit (even) case.. This leads to $Z_{\rm{in}}^{\rm{open}}=-jZ_R\cot\beta x$ and $Z_{\rm{in}}^{\rm{short}}=jZ_R\tan\beta x$. For the even and odd mode the effective impedance is the parallel combination of two  transmission lines of lengths $x_1=\mathcal{\ell}/2$ and $x_2=\left(L-\mathcal{\ell}\right)/2$.
\begin{equation}
\label{even-odd}
\begin{split}
Z_{\rm{tot}}^{e}=&-jZ_R\frac{\cot\left(\beta\mathcal{\ell}/2\right)\cdot\cot\left(\beta\left(L-\mathcal{\ell}\right)/2\right)}{\cot\left(\beta\mathcal{\ell}/2\right)+\cot\left(\beta\left(L-\mathcal{\ell}\right)/2\right)}\\
Z_{\rm{tot}}^{o}=&jZ_R\frac{\tan\left(\beta\mathcal{\ell}/2\right)\cdot\tan\left(\beta\left(L-\mathcal{\ell}\right)/2\right)}{\tan\left(\beta\mathcal{\ell}/2\right)+\tan\left(\beta\left(L-\mathcal{\ell}\right)/2\right)}
\end{split}
\end{equation}
Defining $a(\omega)=\tan\left(\beta\mathcal{\ell}/2\right)$ and $b(\omega)=\tan\left(\beta\left(L-\mathcal{\ell}\right)/2\right)$ and substituting in Eq. \ref{even-odd}, we derive the expressions for the reflections at the port for even and odd mode,
\begin{equation}
\label{gamma_gen}
\begin{split}
\Gamma^e(\omega)=&\frac{-jZ_R/\left(a(\omega)+b(\omega)\right)-Z_0}{-jZ_R/\left(a(\omega)+b(\omega)\right)+Z_0}\\=&\frac{jr+\left(a(\omega)+b(\omega)\right)}{jr-\left(a(\omega)+b(\omega)\right)}\\
\Gamma^o(\omega)=&\frac{jZ_R/\left(a(\omega)^{-1}+b(\omega)^{-1}\right)-Z_0}{jZ_R/\left(a(\omega)^{-1}+b(\omega)^{-1}\right)+Z_0}\\=&\frac{jra(\omega)b(\omega)-\left(a(\omega)+b(\omega)\right)}{jra(\omega)b(\omega)+\left(a(\omega)+b(\omega)\right)}
\end{split}
\end{equation}
where $Z_0$ is the port impedance and $r=Z_R/Z_0$.
The complex amplitude of the scattered waves at port-1 and port-2 are given by the symmetric and antisymmetric superposition of the reflection coefficients $\Gamma^e(\omega)$ and $\Gamma^o(\omega)$ for the even and odd modes respectively:
\begin{equation}
\label{compl_volt}
\begin{split}
K_1=&\frac{1}{2}\Gamma^e(\omega)+\frac{1}{2}\Gamma^o(\omega)\\
K_2=&\frac{1}{2}\Gamma^e(\omega)-\frac{1}{2}\Gamma^o(\omega)
\end{split}
\end{equation}
The ABCD matrix between the two ports attached to the ring resonator is then written as \cite{pozar}:
\begin{equation}
\label{abcd_ring}
\begin{split}
    &\rm{ABCD}_{\rm{ring}}=\begin{pmatrix}
    A_r & B_r \\
    C_r & D_r \\
    \end{pmatrix}\\&=\frac{1}{2\it{K}_{\rm{2}}}\begin{pmatrix}
    \left(1-K_1^2+K_2^2\right) & & Z_0\left(1+K_1^2-K_2^2\right) \\
    \left(1-K_1^2-K_2^2\right)/Z_0 & & \left(1-K_1^2+K_2^2\right) \\
    \end{pmatrix}
\end{split}
\end{equation}
Note that at the special frequency $\omega_{\rm{sp}}=3\omega_0/2$, where $\omega_0$ is the fundamental mode of the ring resonator, the ABCD matrix of the ring is reduced to a simple form:
\begin{equation}
\label{abcd_ring_sp}
\rm{ABCD}_{\rm{sp}}=\begin{pmatrix}
0 & & jZ'(\theta) \\
jY'(\theta) & & 0 \\
\end{pmatrix}
\end{equation}
where, $Z'(\theta)=1/Y'(\theta)=\left(\frac{Z_R}{2}\sin{\frac{3\theta}{2}}\right)$ and $\theta$ is the angular spacing between the qubits in radians.

It is also evident from Eq. \ref{abcd_ring_sp} that choosing $\theta$ at the interval of $\pi/6$ limits the variation of the parameter $Z'$ to a few values: $Z'(2\pi/3)=0$, $Z'(\pi/3)=-Z'(\pi)=Z'_{\rm{max}}$ and $Z'(\pi/6)=Z'(\pi/2)=-Z'(5\pi/6)=(Z'_{\rm{max}}/\sqrt{2})$, where $Z'_{\rm{max}}=Z_R/2$. This leads to the fact that the magnitude of the coupling between connected pairs shows only two different values at the special frequency as mentioned in the main text.

However, in addition to computing the coupling at the special frequency, we are also interested in finding the variation of coupling as we move away from this point. Hence, we continue using the general expression for the ABCD matrix of the ring as given by Eq. \ref{abcd_ring} in the rest of the analysis.
The qubits are introduced in the circuit and are approximated as linear resonators by replacing the Josephson junction with a linear inductor. The qubits are capacitively coupled to the ring resonator with a coupling capacitance $C_g^B$. The ABCD matrix of the coupling capacitance is written as:
\begin{equation}
\label{abcd_cap}
\rm{ABCD}_{\it{C_g^B}}=\begin{pmatrix}
1 & -j/(\omega C_g^B) \\
0 & 1 \\
\end{pmatrix}
\end{equation}
and can be absorbed into a combined ABCD matrix containing the ring resonator and the two coupling capacitors. The combined ABCD matrix is defined as:
\begin{equation}
\label{abcd_tot}
\begin{split}
    &\rm{ABCD}_{comb}=
    \begin{pmatrix}
    \tilde{A} &	\tilde{B} \\
    \tilde{C} &	\tilde{D} \\
    \end{pmatrix}\\&=
    \begin{pmatrix}
    1 & -j/(\omega C_{g}^B) \\
    0 & 1 \\
    \end{pmatrix}
    \begin{pmatrix}
    A_r & B_r \\
    C_r & D_r \\
    \end{pmatrix}
    \begin{pmatrix}
    1 & -j/(\omega C_{g}^B) \\
    0 & 1 \\
\end{pmatrix}
\end{split}
\end{equation}
Substituting for the values of $A_r, B_r, C_r$ and $D_r$ from Eq. \ref{abcd_ring} and simplifying, we get,
\begin{widetext}
\begin{equation}
\label{abcd_vals}
\begin{split}
\tilde{A}=&\frac{2\left(a(\omega)+b(\omega)\right)+C_{g}^BZ_R\omega\left(1-a(\omega)b(\omega)\right)}{C_{g}^BZ_R\omega\left(1+a(\omega)b(\omega)\right)}\\
\tilde{B}=&j\frac{2{C_{g}^B}^2Z_R^2\omega^2a(\omega)b(\omega)-2\left(a(\omega)+b(\omega)\right)^2+2Z_R\omega C_{g}^B\left(a(\omega)+b(\omega)\right)\left(a(\omega)b(\omega)-1\right)}{{C_{g}^B}^2Z_R\omega^2\left(a(\omega)+b(\omega)\right)\left( 1+a(\omega)b(\omega)\right)}\\
\tilde{C}=&\frac{2j}{Z_R}\cdot\frac{a(\omega)+b(\omega)}{1+a(\omega)b(\omega)}\\
\tilde{D}=&\frac{2\left(a(\omega)+b(\omega)\right)+C_{g}^BZ_R\omega\left(1-a(\omega)b(\omega)\right)}{C_{g}^BZ_R\omega\left(1+a(\omega)b(\omega)\right)}
\end{split}
\end{equation}.
\end{widetext}
We then write down the coupled equations of the complete circuit consisting of the pair of linearized qubits and the ring resonator in the frequency domain. The coupling between the pair of qubits is mediated by the ABCD matrix given by Eq. \ref{abcd_vals} for the two port network shown in Fig. \ref{fig:even_odd}(c). The inductance and capacitance of the qubits are given by $L_{Q}=\phi_0^2/E_J$ and $C_{Q}= e^2/2E_C$ respectively, where $E_J$ is the Josephson energy, $E_C$ is the electrostatic charging energy of the oscillators and $\phi_0=\hbar/2e$ is the reduced flux quantum.
\begin{equation}
\label{coupled_eq}
\begin{split}
\frac{i_1}{j\omega C_{Q1}}+j\omega L_{Q1}\left(i_1-i_3\right)=0\\
\frac{i_2}{j\omega C_{Q2}}+j\omega L_{Q2}\left(i_2-i_4\right)=0
\end{split}
\end{equation}
where $i_3$ and $i_4$ are connected by the elements of the ABCD matrix of the black box.
\begin{equation}
\label{cp_abcd}
\begin{pmatrix}
i_3 \\
i_4
\end{pmatrix}=
\frac{1}{\tilde{B}}\begin{pmatrix}
\tilde{D} & \tilde{B}\tilde{C}-\tilde{A}\tilde{D} \\
-1 & \tilde{A} \\
\end{pmatrix}
\begin{pmatrix}
-i_1/(j\omega C_{Q1})\\
-i_2/(j\omega C_{Q2})
\end{pmatrix}
\end{equation}
The coupled equations (\ref{coupled_eq}) are now written as:
\begin{equation}
\label{coupled_eq2}
\begin{split}
\frac{i_1}{C_{Q1}}\tilde{B}+j\omega\frac{L_1}{C_{Q1}}i_1\tilde{D}-&\omega^2L_1i_1\tilde{B}\\&+j\omega\frac{L_1}{C_{Q2}}i_2(\tilde{B}\tilde{C}-\tilde{A}\tilde{D})=0\\
\frac{i_2}{C_{Q2}}\tilde{B}+j\omega\frac{L_2}{C_{Q2}}i_2\tilde{A}-&\omega^2L_2i_2\tilde{B}\\&-j\omega\frac{L_2}{C_{Q1}}i_1=0
\end{split}
\end{equation}
Dividing the equations by $L_1$ and $L_2$ respectively and using qubit frequencies, $\omega_{q,m}=(L_mC_m)^{-1/2}$ we have,
\begin{equation}
\label{coupled_eq3}
\begin{split}
\left(\tilde{B}\omega_{q1}^2+\frac{j\omega}{C_{Q1}}\tilde{D}-\omega^2\tilde{B}\right)i_1+\frac{j\omega}{C_{Q2}}(\tilde{B}\tilde{C}-\tilde{A}\tilde{D})i_2=0\\
-\frac{j\omega}{C_{Q1}}i_1+\left(\tilde{B}\omega_{q2}^2+\frac{j\omega}{C_{Q2}}\tilde{A}-\omega^2\tilde{B}\right)i_2=0
\end{split}
\end{equation}
The eigenmodes $\left\lbrace \omega_\lambda\right\rbrace $ of the system are found by solving the determinant $|\mathcal{M}_{2\times2}|=0$, where $\mathcal{M}_{2\times2}$ is given by:
\begin{equation}
\label{m_el}
\begin{split}
\mathcal{M}_{11}=&\tilde{B}(\omega_{q1}^2-\omega^2)+\frac{j\omega}{C_{Q1}}\tilde{D}\\
\mathcal{M}_{12}=&\frac{j\omega}{C_{Q2}}\left(\tilde{B}\tilde{C}-\tilde{A}\tilde{D}\right)\\
\mathcal{M}_{21}=&-\frac{j\omega}{C_{Q1}}\\
\mathcal{M}_{22}=&\tilde{B}(\omega_{q2}^2-\omega^2)+\frac{j\omega}{C_{Q2}}\tilde{A}
\end{split}
\end{equation}

When the two qubits are set identical, the splitting between the two eigen frequencies corresponding to the qubit modes is given by $\Delta=2J_{ij}$, where $J_{ij}$ is the inter-qubit coupling at the mean of the two frequencies. 


\section{Scaling up to more qubits}
To scale up, one of the proposed schemes in the main text is the use of a larger ring having the fundamental mode at 1 GHz and qubits placed at $10^o$ angular spacing. The special operating point is now at the mean of the third and fourth harmonics, i.e $\tilde{\omega}_Q=(\omega_R^3+\omega_R^4)/2=4.5$ GHz. We use the same treatment as mentioned above to generate the inter-qubit coupling as a function of frequency and angle. 
Any qubit (e.g. one highlighted in yellow, in Fig 4(a) of the main text) is connected to 27 other qubits in the ring. In this longer ring, qubits spaced by $40^o$ to each other show negligible coupling. The remaining qubits are connected with two slightly different values of coupling at $\tilde{\omega}_Q$, shown as red and blue  dots in the figure representing the stronger and weaker coupling respectively. However, the longer ring resonator results in a reduced window for the choice of qubit frequencies as the coupling values deviate a lot faster as one moves from the special frequency. This is due to a smaller spacing between the successive harmonics of the ring resonator, leading to  stronger variation of coupling with frequency.

The other approach involves multiple ring cavities with each pair of rings connected via a $\lambda/2$ resonator resonating at the fundamental mode of the ring cavity. The characteristic impedance of the $\lambda/2$ section is chosen to be $Z_C=Z_R/2$ so that the impedance is matched and there is no reflection at the junction. We first investigate the case of two connected ring resonators using the ABCD matrices of individual sections. The combined ABCD matrix between the two qubits connected at arbitrary locations of two different rings can be written as:
\begin{widetext}
\begin{equation}
\label{abcd_two_ring}
\rm{ABCD}_{ext}=
\begin{pmatrix}
1 & -j/(\omega C_{g}^B) \\
0 & 1 \\
\end{pmatrix}
\begin{pmatrix}
A_{r1} & B_{r1} \\
C_{r1}  & D_{r1}  \\
\end{pmatrix}
\begin{pmatrix}
A_{\lambda/2} & B_{\lambda/2}  \\
C_{\lambda/2}  & D_{\lambda/2}  \\
\end{pmatrix}
\begin{pmatrix}
A_{r2}  & B_{r2} \\
C_{r2}  & D_{r2} \\
\end{pmatrix}
\begin{pmatrix}
1 & -j/(\omega C_{g}^B) \\
0 & 1 \\
\end{pmatrix}
\end{equation}
\end{widetext}
where the ABCD matrix with subscript $r1$($r2$) corresponds to the two-port network between the qubit and the point where the $\lambda/2$ section is connected to the first (second) ring resonator. The ABCD matrix with subscript $\lambda/2$ corresponds to the section between the two ends of the $\lambda/2$ resonator and is given by:
\begin{equation}
\label{abcd_lambdaby2}
\begin{pmatrix}
A_{\lambda/2} & B_{\lambda/2}  \\
C_{\lambda/2}  & D_{\lambda/2}  \\
\end{pmatrix}
=\begin{pmatrix}
\cos(\beta L_{\lambda/2}) & j Z_C \sin(\beta L_{\lambda/2})  \\
(j/Z_C) \sin(\beta L_{\lambda/2})  & \cos(\beta L_{\lambda/2})  \\
\end{pmatrix}
\end{equation}

where $L_{\lambda/2}$ is the length of the $\lambda/2$ section.

The coupled equations of motion for the two qubits are written in terms of $\mathcal{M}$ given by Eq. \ref{m_el} where we substitute the expressions for $\tilde{A}$, $\tilde{B}$, $\tilde{C}$ and $\tilde{D}$ from Eq. \ref{abcd_two_ring}. Finally, the inter-qubit coupling is computed by bringing the qubits on resonance and measuring the splitting of the qubit normal modes  by solving $|\mathcal{M}_{2\times2}|=0$.

We can use this approach to analyse the full network shown in Fig. 4(c) of the main text. The central ring resonator is connected to six outer ring resonators by $\lambda/2$ sections. The connections are made to the central ring at $60^o$  angular separations. The remaining positions  on the central ring resonator are connected to qubits and six qubits are placed at each of the outer ring resonators uniformly with the first qubit anchored at the $30^o$ angular position from the connecting point. We highlight one qubit in one of the outer rings with yellow in Fig. 4(c) of the main text and show its connectivity to all other qubits in the network. Each qubit is connected to three other qubits sharing the same ring at the locations marked with red dots. Additionally, it is connected to every qubit in the central ring as well as three other outer rings (marked with blue dots). However, all qubits (faded out in the diagram) placed in the two remaining outer rings  are completely decoupled from the highlighted qubit. This happens due to the same destructive interference effect discussed earlier, as those two rings are connected to the central ring at $120^o$  angles with respect to the one containing the highlighted qubit. The qubits are operated around the mean frequency $\tilde{\omega}_Q=(\omega_R^0+\omega_R^1)/2$ of the first two modes of the ring resonators. Remarkably, at the special frequency $\tilde{\omega}_Q$, the inter-qubit couplings share only two different values as before, a higher value for qubits within the same ring and a lower value for qubits from different rings, distinguished by the red and blue dots in the figure.

\begin{figure*}[!t]
    \centering
    \includegraphics[width=0.95 \textwidth]{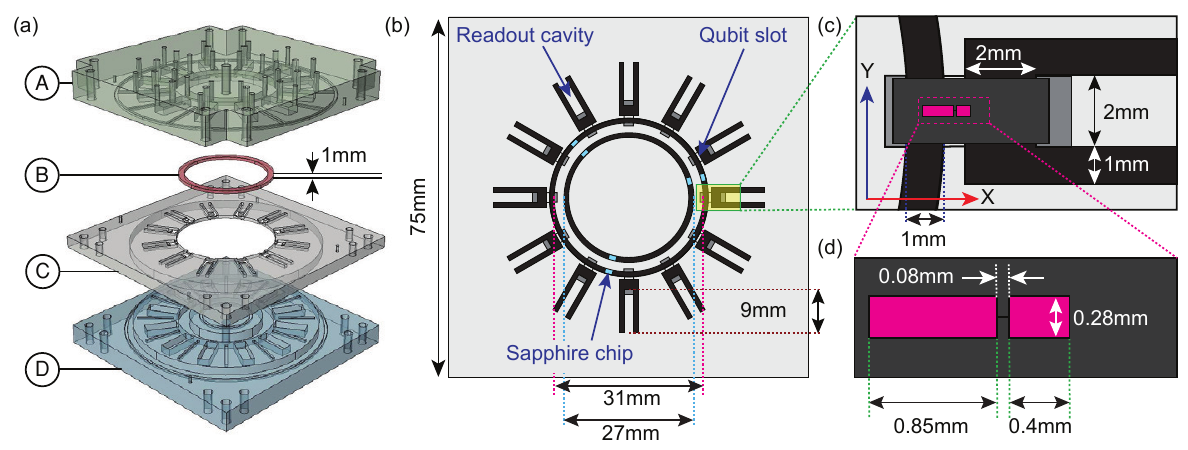} 
    \vspace{-10pt}
    \caption{Design and assembly of the 3D ring cavity set-up. (a) The cavity is machined in four parts labelled (A), (B), (C) and (D). Parts (A) and (D)  contain the top and bottom parts, respectively, of the ring cavity and the readout cavities. Part (B) is the central conductor of the ring cavity. Part (C) contains the central conductor of the readout cavities and the slots for placing the qubits. The SMA connectors (not shown) for the readout cavities are attached to part (A) from the top and bring in the readout and qubit excitation. (b) Top view of the ring cavity setup after assembling parts (B), (C) and (D).  The dimensions of the bus and readout cavities are shown. Three sapphire chips (shown in cyan) hold the central conductor of the bus cavity in position. (c) Magnified view of a qubit chip placed in its slot shared between the bus and readout cavities. (d) Magnified view of the transmon qubit design showing the capacitor pad dimensions used in the experiment.} 
    \label{fig:design}
\end{figure*}

\begin{table*}[t]
	\centering
	\begin{tabular}{c l c c c c c} 
		\hline
		\hline
		\\[-2ex]
		& Measured parameters & Qubit 1 & Qubit 3 & Qubit 9 & Qubit 10   &    \\
		&  & (Q1) & (Q3) & (Q9) & (Q10)   &    \\
		\hline
		\hline \\[-1.5ex]
		& Qubit frequency, ${\omega_q}/{2\pi}$ (GHz) & 4.6376 & 4.5932 & 	4.6566	& 4.7488 &\\
		& Anharmonicity, ${\delta_q}/{2\pi}$ (GHz) & -0.318 & -0.306 &-0.309 & -0.308 &\\ [0.5ex]
		\hline \\[-1.5ex]
		& Readout resonator frequency, ${\omega_R}/{2\pi}$ (GHz) & 7.5500 & 7.5650 & 7.4744 & 7.5095 &\\
		& Readout resonator linewidth, ${\kappa_R}/{2\pi}$ (MHz) & 3.27 & 4.01 & 4.78 & 2.04 &\\
		& Qubit-readout coupling, ${g}/{2\pi}$ (MHz) & 71 & 55 & 73 & 79 &\\ [0.5ex]
		\hline \\[-1.5ex]
		& Relaxation time, $T_1$ ($\mu$s) & 41 & 31 & 37 & 35 &\\
		& Ramsey time, $T_2^R$ ($\mu$s) & 3.4 & 2.6* & 6.0* & 3.6* &\\
		& Hahn echo time, $T_2^E$ ($\mu$s) & 8 & 24 & 20 & 29 &\\ [0.5ex]
		\hline
		\hline
	\end{tabular}
	\caption{Measured device parameters and coherence times of the four transmons used in the experiment. The qubit-readout cavity coupling is estimated by measuring the shift in the cavity frequency when probed at low power and high power. Ramsey fringes of the qubits marked with asterisk show beating leading to deterioration of Ramsey time which improves significantly with an echo sequence.} 
	\label{table:deviceparameters}
\end{table*}

\section{Device design and parameters}
Our experimental design consists of a central circular bus resonator with twelve slots distributed along its perimeter to place the qubits with an angular spacing of $30^o$. 
Each qubit slot is attached to a dedicated $\lambda/4$ readout resonator extending radially outward as shown in Fig. \ref{fig:design}(b). Both the bus resonator and the readout resonators are realized in rectangular coaxial transmission line geometry. Coupling ports for readout and qubit drive are introduced along the third dimension, offering in-plane scalability and convenient 3D integration for control lines.

The ring cavity was designed in four parts and individual parts are machined from aluminum. As shown in Fig. \ref{fig:design}(a), the top and bottom-most pieces contain the top and bottom halves of the ring and readout cavities. The ring shaped central conductor is placed in the bottom piece and rests on three sapphire holder chips as indicated in the figure. The sapphire chips are glued to the bottom piece and the ring is glued to the sapphire chips using a small amount of Stycast 2850FT. It  is carefully placed in position under an optical microscope so that it rests symmetrically. The central piece contains the central conductors of all the twelve readout cavities. They are designed to have nominally identical lengths and hence identical resonant frequencies. The actual machined piece has some unavoidable variations in the dimensions due to machining tolerances. The dimensions of the ring and the readout cavities are indicated in Fig. \ref{fig:design}(b). Once the ring is in place, the central piece is bolted to the bottom piece with four screws. The qubit chips can now be placed in the respective slots as indicated in Fig. \ref{fig:design}(c). This operation is also carried out using an optical microscope for precise placement of the chips and we use N-grease to keep the chips in place. After the qubit chips are placed, the top piece is bolted to the rest of the assembly with nine screws. As can be seen in Fig. \ref{fig:design}(a), several grooves are provided for indium sealing to create appropriate microwave isolation for all readout cavities and the central ring cavity as well.

The measured device parameters and coherence numbers are listed in Table \ref{table:deviceparameters}.

\section{Tuning qubit-cavity couplings}

\begin{figure}[!b]
	\centering
	\includegraphics[width=0.48\textwidth]{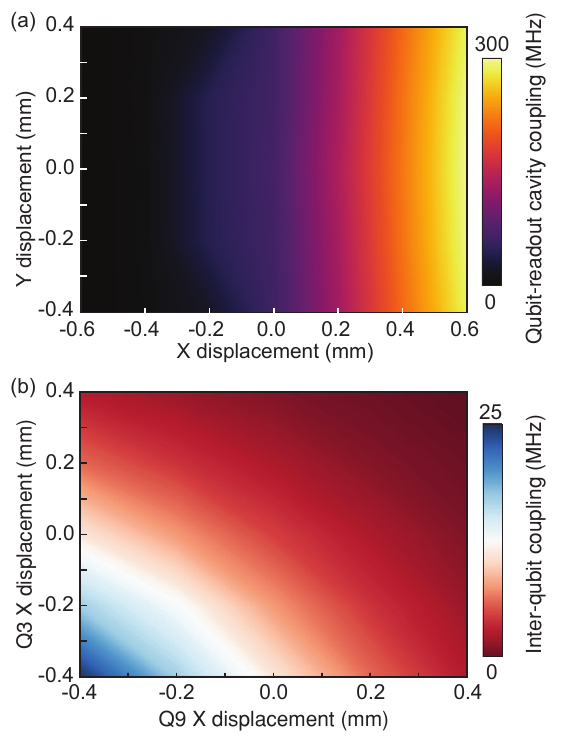} 
	\vspace{-15pt}
	\caption{Tolerance in qubit placement estimated from finite element simulation. (a) Variation of qubit-readout coupling as a function of qubit displacement in the X direction (into and out of the readout cavity) and Y direction (lateral displacement). We find that the coupling is weakly dependent on lateral displacement but it is more sensitive to the in and out displacement of the qubit. (b) Inter-qubit coupling between the pair Q3-Q9 placed at $180^o$ as a function of X displacement of both the qubits. The displacement determines the coupling with the bus cavity and hence the inter-qubit coupling. The coupling is sensitive to qubit placement requiring precise placement of the qubits in their respective slots.}
	\label{fig:tolerance}
\end{figure}

\begin{table*}[t]
	\centering
	\begin{tabular}{c l c c c c c c c c c c c c c c} 
		\hline
		\hline
		\\[-2ex]
		 & Qubit pair  && Q9-Q10 && Q1-Q3 && Q1-Q10 && Q1-Q9 && Q3-Q10 &&  Q3-Q9   &\\
		\hline
		\hline \\[-1.5ex]
	    & Relative angular positions && $30^o$ && $60^o$  && $90^o$ && $120^o$  && $150^o$ && $180^o$   & \\  [0.5ex]
	    & Measured cross-Kerr shift (kHz) && -102 && -140 && -28 && -5  && -104 && -146   & \\  [0.5ex]
	    & Estimated inter-qubit coupling (MHz) && 3.45 && 4.57 && 2.40 && 0.05  && -3.58 && -4.74   & \\  [0.5ex]
		\hline
		\hline
	\end{tabular}
	\caption{Cross-Kerr shift between all qubit pairs measured in the conditional Ramsey experiment and the estimated inter-qubit coupling $J_{ij}$.}
	\label{table:jazz}
\end{table*}

The capacitor pads of the qubits are designed based on finite element simulation of the complete qubit-cavity system to yield desired qubit-readout coupling and inter-qubit coupling. The qubit junction sits on the bridge between the readout cavity and the bus cavity and the dimension of the capacitor pad extending into respective cavities determines the magnitudes of the couplings. 
In the finite element simulation we eliminate the Josephson junction and introduce a port across the junction terminals and extract the full scattering matrix of the system.
To estimate the qubit-readout cavity coupling we terminate the read cavity port with a $50 \Omega$ load and compute the resulting Purcell $T_1$ decay time as a function of the detuning between the qubit and the readout cavity. We then fit this to the Purcell $T_1$ formula to extract the coupling.

We use an avoided crossing simulation to numerically extract the inter-qubit coupling. We approximate the qubits as harmonic oscillators and introduce two linear inductors to the two-port circuit represented by the scattering matrix. Finally, we bring the two oscillators on resonance by tuning the inductor values to produce an avoided crossing. The inter-qubit coupling is then estimated from the normal mode splitting. 

We also analyze the tolerance of the coupling of qubit to readout cavity and the inter-qubit coupling to small deviations in placement of the qubits in their slots. In Fig. \ref{fig:tolerance}(a) we plot the dependence of the qubit-readout coupling as a function of X and Y displacement (see Fig. \ref{fig:design}(c)). X displacement refers to the direction parallel to the central conductor of the readout resonator; Y displacement is in the direction orthogonal to X. We observe that the coupling is relatively insensitive to Y displacement but it depends strongly on the displacement in X direction. Next we study the dependence of the inter-qubit coupling on the X displacement of the qubits and the results are shown in Fig. \ref{fig:tolerance}(b). The data indicates that chips have to be placed precisely to prevent large deviations in inter-qubit couplings.  We use an optical microscope to place the qubits precisely in their slots.

\section{Estimation of the inter-qubit coupling}
The effective Hamiltonian of the coupled four transmon system is written as:
\begin{equation}
    \label{four_qubit}
    \begin{split}
    \hat{\mathcal{H}}_0/\hbar=&\sum_{i=1}^{N}(\omega_i\hat{a}^\dagger_i\hat{a}_i+\frac{\delta_i}{2}\hat{a}^\dagger_i\hat{a}_i(\hat{a}^\dagger_i\hat{a}_i-\mathbf{I}))\\&+\sum_{i<j,j=1}^{N}J_{\rm{ij}}(\hat{a}^\dagger_i\hat{a}_j+\hat{a}^\dagger_j\hat{a}_i)
    \end{split}
\end{equation}
where, $a_i$ and $a_i^\dagger$ are the annihilation and creation operators of the $k^{th}$ transmon which is modeled as Duffing oscillator, $\tilde{\omega}_i$ and $\delta_i$ are the lowest transition frequency and the anharmonicity of the respective ladders. $J_{ij}$ is the exchange coupling between the $i^{th}$ and $j^{th}$ transmons. We have neglected the frequency dependence of the exchange coupling in this calculation.

The inter-qubit coupling terms hybridize the levels and the eigen-modes of the coupled four-transmon system are obtained by diagonalizing the Hamiltonian in Eq. \ref{four_qubit}. In the experiment we extract the effective low energy levels of the full Hamiltonian in the diagonal (dressed) basis. The problem then reduces to numerically finding the best set of coupling parameters $J_{ij}$ in Eq. \ref{four_qubit} with the measured values of $\tilde{\omega}_i$ and $\delta_i$ that would produce the experimentally observed low energy eigen spectra.

The cross-Kerr shift is a manifestation of the interaction among the doubly excited levels e.g. $|0002\rangle$, $|0020\rangle$, $|0011\rangle$, $|1001\rangle$ etc. that depends on the exchange coupling $J_{ij}$ and can be measured by a set of conditional Ramsey sequences. The vacuum Rabi splitting experiment is an alternate method to extract the exchange coupling values. The splitting depends on the interaction between the single excitation levels of all the qubits. 

While numerically finding the actual inter-qubit coupling using Eq. \ref{four_qubit}, we use 8 levels for each transmon and truncate the total number of excitations in the system at 8. We have also restricted the search space by assigning appropriate polarity of the coupling which is negative for angles between $120^o$ and $240^o$ and positive otherwise.
The estimated coupling from the conditional Ramsey experiment between all pairs are listed in Table \ref{table:jazz}.

\begin{figure}[!t]
	\centering
	\includegraphics[width=0.48\textwidth]{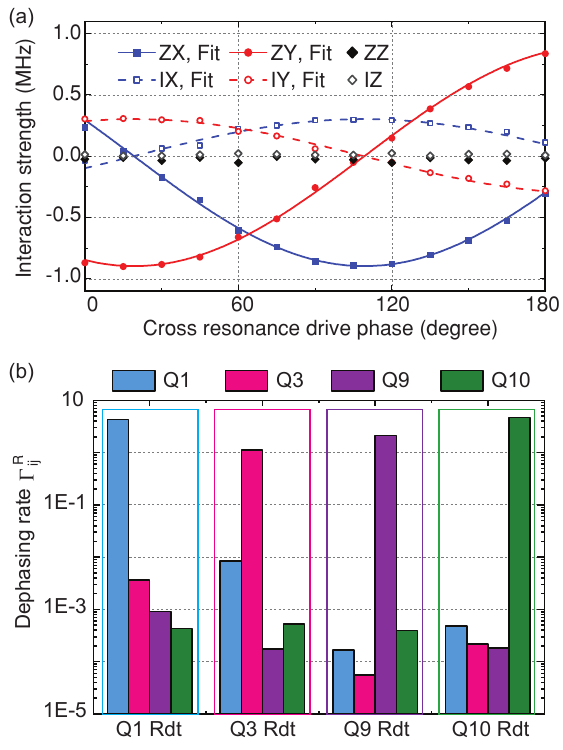} 
	\vspace{-10pt}
	\caption{(a) Characterizing the microwave cross-talk around qubit frequencies. A Hamiltonian tomography experiment is performed under the application of cross resonance drive to quantify the effective interaction terms. The corresponding interaction terms for the pair Q1-Q10 are plotted as a function of the phase of the cross resonance drive applied on Q10. We estimate the classical cross-talk  as defined in Eq. \ref{drive} to be $m=0.1\%$ for this pair. (b) Measurement induced dephasing rate on the targeted qubit as well as the un-targeted qubit due to microwave cross-talk at readout frequencies. Transparent larger box indicates the targeted qubit where we apply the measurement tone whereas the solid boxes refer to the qubit where we perform the echo sequence. We observe a negligible effect on the un-targeted qubits measuring at least three orders of magnitude lower dephasing rate due to spurious cross-talk except for the Q1-Q3 pair, where the difference is just below 1\%.}
	\label{fig:ct}
\end{figure}

\begin{table*}[t]
	\centering
	\begin{tabular}{c l c c c c c c c c c} 
		\hline
		\hline
		\\[-2ex]
		 & CR pair  && Q9-Q10 && Q1-Q3 && Q1-Q10 &&  Q3-Q9   &\\
		\hline
		\hline \\[-1.5ex]
	    & Relative angular positions && $30^o$ && $60^o$  && $90^o$ && $180^o$   & \\  [0.5ex]
	    & CR drive amplitude, $\Omega$ (MHz) && 28.9 && 14.7  && 27.5 && 21.4   & \\  [0.5ex]
		\hline \\[-1.5ex]
				& CR cross-talk ($m$) && 0.011 &&0.008 &&0.001 && 0.008 &\\ [0.5ex]
		\hline
		\hline
	\end{tabular}
	\caption{Measured classical microwave cross-talk between the cavities around readout frequencies and qubit frequencies. CR is not performed between $120^o$ pair due to negligible coupling and small detuning ($<20$ MHz) between the qubits. Also, the pair at $150^o$ has a resonance between the $\left(|0\rangle\leftrightarrow|1\rangle\right)$ transition of one qubit and the  $\left(|0\rangle\leftrightarrow|2\rangle\right)/2$ transition of the other.}
	\label{table:cross_talk}
\end{table*}

\section{Qubit drive cross-talk}
Classical cross-talk due to microwave leakage at qubit frequencies to un-targeted qubits may lead to unwanted evolution of the system. We have used cross-resonance(CR)\cite{cr_chow} interaction in our multi-qubit architecture to characterize the classical cross-talk. The CR Hamiltonian is given by,
\begin{equation}
    \label{cr_ham}
    \begin{split}
    H_{\rm{CR}}/\hbar=&\sum_{k=1, 2}{\left(\tilde{\omega}_k b_k^\dagger b_k+\frac{\delta_k}{2}b_k^\dagger b_k\left(b_k^\dagger b_k-\mathbf{I}\right)\right)}\\&+ J\left(b_1^\dagger b_2+b_2^\dagger b_1\right)+H_d
    \end{split}
\end{equation}
where, $b_k$ and $b_k^\dagger$ are the annihilation and creation operator of the $k^{th}$ transmon, $\tilde{\omega}_k$ and $\delta_k$ are the lowest transition frequency and the anharmonicity of the respective ladders. $J$ is the exchange coupling strength between the two transmons and the drive term $H_d$ is given by,
\begin{equation}
    \label{drive}
    \begin{split}
    H_d=& \Omega \cos\left(\tilde{\omega}_2t+\varphi_0\right)\left(b_1^\dagger + b_1\right)\\&+m\Omega \cos\left(\tilde{\omega}_2t+\varphi_{\rm{CT}}\right)\left(b_2^\dagger + b_2\right)
    \end{split}
\end{equation}
where the second term is due to microwave cross-talk directly driving the second qubit. The resulting ZX and IX interaction terms can be extracted by an effective Hamiltonian theory and when the drive phase $\varphi_0$ is set to zero, the amplitudes of the two terms are given by\cite{effective_hamiltonian_cr},
\begin{equation}
    \label{zx_term}
    \begin{split}
    A_{\rm{ZX}}=&-\frac{J\Omega}{\Delta}\left(\frac{\delta_1}{\delta_1+\Delta}\right)\\&+ \frac{J\Omega^3\delta_1^2\left(3\delta_1^3+11\delta_1^2\Delta+15\delta_1\Delta^2+9\Delta^3\right)}{4\Delta^3\left(\delta_1+\Delta\right)^3\left(\delta_1+2\Delta\right)\left(3\delta_1+2\Delta\right)}
    \end{split}
\end{equation}
\begin{equation}
    \label{ix_term}
    A_{\rm{IX}}=-\frac{J\Omega}{\delta_1+\Delta}+ \frac{J\Omega^3\delta_1\Delta}{\left(\delta_1+\Delta\right)^3\left(\delta_1+2\Delta\right)\left(3\delta_1+2\Delta\right)}
\end{equation}
where $\Delta$ is the detuning between the two transmons.

In the experiment, we apply the CR drive and perform Hamiltonian tomography\cite{hamiltonian_tomography_ibm} to extract the effective interaction terms emerging from the drive. For a fixed amplitude of the cross-resonance drive we vary the phase of the CR drive and plot the interaction terms as a function of drive phase. The ZZ and IZ terms are independent of the drive phase, whereas the ZX, ZY, IX and IY terms oscillate periodically (See Fig. \ref{fig:ct}(a)). We simultaneously fit the oscillations to sine functions.

We compute $\Omega$, the applied CR drive amplitude from the amplitude of the ZX oscillation, $A_{\rm{ZX}}$ using Eq. \ref{zx_term} and plug it to Eq. \ref{ix_term} to evaluate $A_{\rm{IX}}$, the amplitude of IX term arising from CR interaction, in the absence of any cross-talk. Next we measure the phase difference between the ZX and IX oscillations,  $\varphi_{\rm{CT}}=\varphi_{\rm{ZX}}-\varphi_{\rm{IX}}$ and finally compute the cross-talk factor  $m=\Omega_{\rm{CT}}/\Omega$, where $\Omega_{\rm{CT}}$ is expressed as,
\begin{equation}
    \Omega_{\rm{CT}}=\sqrt{\left(A_{\rm{IX}}\right)^2+2\cos(\varphi_{\rm{CT}})A^M_{\rm{IX}}A_{\rm{IX}}+\left(A^M_{\rm{IX}}\right)^2}
\end{equation}
where $A^M_{\rm{IX}}$ is the experimentally measured amplitude of the IX oscillations. 

We estimate classical cross talk for four qubit pairs in our system at $30^o$, $60^o$, $90^o$  and $180^o$. The corresponding values are listed in Table. \ref{table:cross_talk}. The other two pairs are not compatible for a cross resonance experiment as the pair at $120^o$ has negligible coupling and the pair at $150^o$ has a detuning such that the $\left(|0\rangle\leftrightarrow|1\rangle\right)$ transition of the target qubit is resonant with $\left(|0\rangle\leftrightarrow|2\rangle\right)/2$ transition of the control qubit. However, we expect the cross-talk to be similar to the other values measured.

Our estimate of cross-talk is in congruence with a previous result\cite{cr_dimon} showing significantly lower cross-talk in 3D cQED architecture compared to the existing 2D designs\cite{effective_hamiltonian_cr} due to superior microwave isolation between the cavities.

\section{Readout drive cross-talk}
We characterize the effect of measurement cross-talk on the un-targeted qubits while performing readout\cite{mult_rdt}. To evaluate the effect of microwave cross-talk  we perform a Hahn echo experiment and measure the dephasing rate of the un-targeted qubit ($Q_i$) while turning on a continuous calibrated readout tone on the targeted qubit ($Q_j$). We then compute the excess dephasing rate per photon using the equation:
\begin{equation}
    \Gamma_{ij}^R=\frac{1}{\bar{n}}\left(\frac{1}{\tau_i^j}-\frac{1}{\tau_i^0}\right)
\end{equation}
    where, $\bar{n}$ is the average photon number used for the calibrated tone on the targeted qubit. Here $\tau_i^0$ and $\tau_i^j$ are Hahn echo times of $Q_i$ with and without the additional tone on $Q_j$ respectively. We used average photon numbers $\bar{n}=$ 13, 19, 11 and 12 for qubits Q1, Q3, Q9 and Q10 to optimize the readout fidelity for each qubit. In Fig. \ref{fig:ct}(b), we show the effect of cross-talk by plotting the values of $\Gamma_{ij}^R$. When compared to the targeted qubit, we observe more than three orders of magnitude smaller dephasing rate per photon for the un-targeted qubits. The only exception is the Q1-Q3 pair, which has a slightly higher dephasing rate.
}

%


\cleardoublepage

	\setcounter{figure}{0}
	\setcounter{table}{0}
	\setcounter{equation}{0}
	\setcounter{section}{0}
	\newpage
	\global\long\def\theequation{S\arabic{equation}}
	\global\long\def\thefigure{S\arabic{figure}}
	\global\long\def\thetable{S\arabic{table}}
	\global\long\def\thesection{S\arabic{section}}
	\onecolumngrid
	\begin{center}
		{\bf \large Supplementary Information: Long-range connectivity in a superconducting quantum processor using a ring resonator}
	\end{center}

	\begin{center}	
		{Sumeru Hazra$^{1}$, Anirban Bhattacharjee$^{1}$, Madhavi Chand$^{1}$, Kishor V. Salunkhe$^{1}$, Sriram Gopalakrishnan$^{1,2}$,\\
		Meghan P. Patankar$^{1}$ and R. Vijay$^{1}$}\\
		$^{1}$\textit{Department of Condensed Matter Physics and Materials Science, \\ Tata Institute of Fundamental Research, Homi Bhabha Road, Mumbai 400005, India and\\ $^{2}$Current affiliation: Institute of Quantum Computing, \\ University of Waterloo, Waterloo, Ontario N2L 3G1, Canada}
	\end{center}

\section{Raw data for estimating inter-qubit coupling}
\begin{figure}[b]
	\centering
	\includegraphics[width=\textwidth]{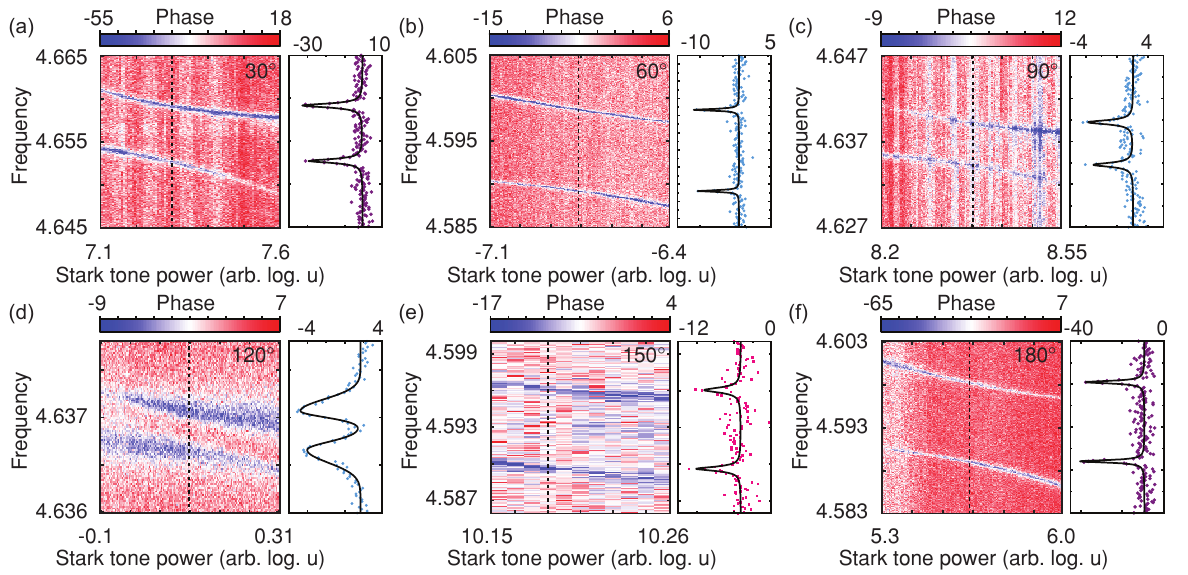} 
	\vspace{-5pt}
	\caption{(a-f) Avoided crossing data between all qubit pairs  for 30, 60, 90, 120, 150 and 180 degrees. The right panels show vertical line cuts when the qubits are maximally hybridized. They are fitted with a double Lorentzian function except the one in D which is fitted with an asymmetric bi-Gaussian double peak function. With Q10 being the highest frequency qubit, we need to apply large Stark power to achieve the desired shift in frequency. That is why the spectra in Fig. {\bf{a}},{\bf{c}} and especially {\bf{d}} look noisier compared to the others.}
	\label{fig:exp_coupling_anticrossing}
\end{figure}

The avoided crossing observed in the spectroscopy on all six qubit pairs are plotted in  Fig. \ref{fig:exp_coupling_anticrossing}. We also plot the line cuts when the two peaks are at the nearest. We fit the spectroscopy data with a double Lorentzian function and extract the vacuum Rabi splitting from the fitting parameters. However, the peaks were not well separated in the case of the $120^o$ pair. We have used an asymmetric bi-Gaussian function to estimate the peak frequencies for this particular set.

\begin{figure}[!t]
	\centering
	\includegraphics[width=\textwidth]{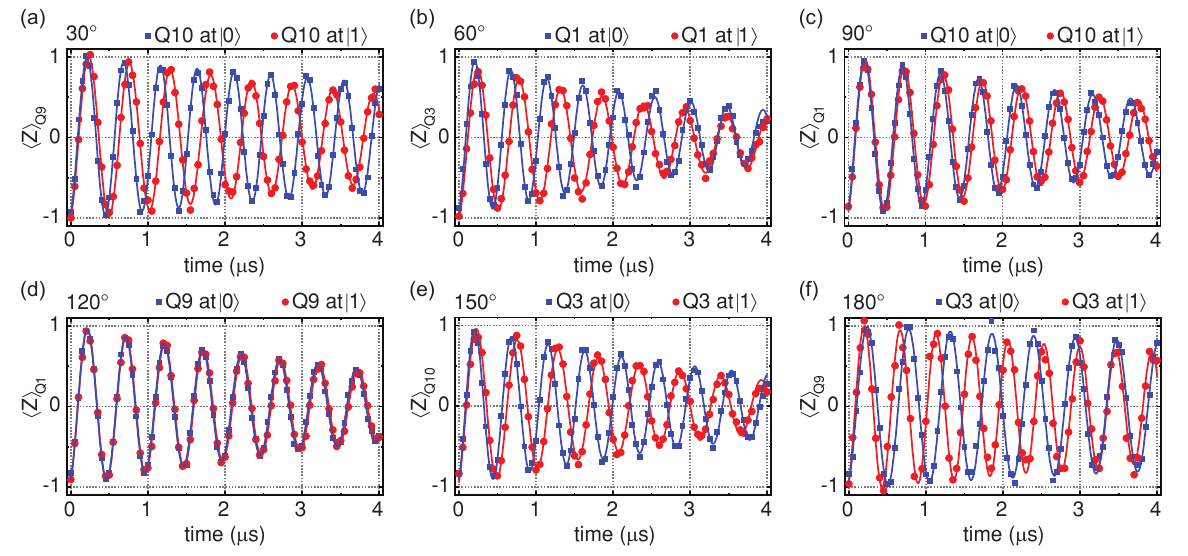} 
	\vspace{-10pt}
	\caption{(a-f) Estimating inter-qubit coupling between all pairs for $30^o$, $60^o$, $90^o$, $120^o$, $150^o$ and $180^o$ by  measuring the cross-Kerr shift. A conditional Ramsey interferometry sequence is performed on each pair to measure the cross-Kerr shift. The pairwise inter-qubit exchange coupling is then estimated from the measured cross-Kerr shifts.}
	\label{fig:exp_coupling_jazz}
\end{figure}

As an independent measurement of the inter-qubit coupling we perform a conditional Ramsey experiment to quantify the effective cross-Kerr interaction between all pairs. We implement a Ramsey sequence on each qubit to measure its frequency with the other qubit of that pair in the ground and excited states respectively (See Fig \ref{fig:exp_coupling_jazz}). The difference gives us the cross-Kerr shift. Finally, we numerically estimate the transverse coupling strengths that are most likely to have produced the experimentally measured cross-Kerr shifts on all the pairs. 

It is to be noted that the cross-Kerr shift results from the interaction of doubly excited levels of the multi-qubit system and depends on the coupling strength between them. Therefore the shift depends on the inter-qubit coupling strength both at the $|g\rangle$ to $|e\rangle$ transition, as well as the $|e\rangle$ to $|f\rangle$ transition frequencies of the qubits. In our case, the inter-qubit coupling is a slowly varying function of frequency around the special frequency and hence we don't expect a big difference between the two techniques. However, if the coupling is strongly frequency dependent, the two methods will, in general,  yield different results.

\clearpage

\end{document}